\documentclass[sigconf]{acmart}

\usepackage[utf8]{inputenc}
\usepackage{tabularx}
\usepackage{booktabs}
\usepackage{soul}
\usepackage{xcolor}
\usepackage{xspace}
\usepackage{graphicx}
\usepackage[caption=false]{subfig}  
\usepackage[export]{adjustbox}
\usepackage{amsmath}
\usepackage{textcomp}
\usepackage{listings}
\usepackage{cleveref}
\usepackage{algorithm}
\usepackage[noend]{algpseudocode}
\usepackage{balance}
\usepackage{enumitem}
\usepackage{multirow}
\usepackage{colortbl}
\usepackage{balance}
\usepackage{threeparttable}
\usepackage{multicol}

\definecolor{codegreen}{rgb}{0,0.6,0}
\definecolor{codegray}{rgb}{0.5,0.5,0.5}
\definecolor{codepurple}{rgb}{0.58,0,0.82}
\definecolor{backcolour}{rgb}{0.95,0.95,0.92}
\definecolor{gray}{gray}{0.9}
\definecolor{APA_stats}{RGB}{100, 100, 120}

\newcommand{\CR}{Code Readability\xspace}
\newcommand{\ie}{\emph{i.e.,}\xspace}
\newcommand{\eg}{\emph{e.g.,}\xspace}

\usepackage{enumitem}
\usepackage{tcolorbox}

\newenvironment{change}{}{\xspace}

\newcounter{observation}
\newcommand{\observation}[1]{\refstepcounter{observation}
	\begin{center}
		\framebox{
			\begin{minipage}{0.9\columnwidth}
				{} \textit{#1}
			\end{minipage}
		}
	\end{center}
}
\AtBeginDocument{%
  }

\setcopyright{acmcopyright}
\copyrightyear{2023}
\acmYear{2023}
\acmDOI{XXXXXXX.XXXXXXX}

\acmConference[ICPC 2024]{The 32nd IEEE/ACM International Conference on Program Comprehension}{April 15--16,
  2024}{Lisbon, Portugal}
\acmPrice{15.00}
\acmISBN{978-1-4503-XXXX-X/18/06}

\begin{document}

\title[Reassessing Code Readability Models]{Reassessing Java Code Readability Models with a Human-Centered Approach}


\author{Agnia Sergeyuk}
\email{agnia.sergeyuk@jetbrains.com}
\affiliation{%
  \institution{JetBrains Research}
  \city{Belgrade}
  \country{Serbia}
}

\author{Olga Lvova}
\email{olga.lvova@jetbrains.com}
\affiliation{%
  \institution{JetBrains}
  \city{Yerevan}
  \country{Armenia}
}

\author{Sergey Titov}
\email{sergey.titov@jetbrains.com}
\affiliation{%
  \institution{JetBrains Research}
  \city{Paphos}
  \country{Cyprus}
}

\author{Anastasiia Serova}
\email{anastasiia.serova@jetbrains.com}
\affiliation{%
  \institution{JetBrains}
  \city{Paphos}
  \country{Cyprus}
}

\author{Farid Bagirov}
\email{farid.bagirov@jetbrains.com}
\affiliation{%
  \institution{JetBrains Research}
  \city{Paphos}
  \country{Cyprus}
}

\author{Evgeniia Kirillova}
\email{evgeniia.kirillova@jetbrains.com}
\affiliation{%
  \institution{JetBrains Research}
  \city{Munich}
  \country{Germany}
}

\author{Timofey Bryksin}
\email{timofey.bryksin@jetbrains.com}
\affiliation{%
  \institution{JetBrains Research}
  \city{Limassol}
  \country{Cyprus}
}

\renewcommand{\shortauthors}{Sergeyuk et al.}


\begin{abstract}
To ensure that Large Language Models (LLMs) effectively support user productivity, they need to be adjusted. Existing Code Readability (CR) models can guide this alignment. However, there are concerns about their relevance in modern software engineering since they often miss the developers’ notion of readability and rely on outdated code. This research assesses existing Java CR models for LLM adjustments, measuring the correlation between their and developers’ evaluations of AI-generated Java code. Using the Repertory Grid Technique with 15 developers, we identified 12 key code aspects influencing CR that were consequently assessed by 390 programmers when labeling 120 AI-generated snippets. Our findings indicate that when AI generates concise and executable code, it's often considered readable by CR models and developers. However, a limited correlation between these evaluations underscores the importance of future research on learning objectives for adjusting LLMs and on the aspects influencing CR evaluations included in predictive models.
\end{abstract}


\maketitle

\section{INTRODUCTION} \label{sec:introduction}

Recently, the field of Large Language Models (LLMs) has been rapidly developing, enabling a wide range of applications in various domains, including software development (for review, see~\cite{kaddour2023challenges}). In programming, code-fluent LLMs are used as coding assistants --- AI is mainly used for code generation and completion. It also can suggest code refactoring, commit messages, code explanations, and has many other practical programming-related applications, \eg Copilot,\footnote{Copilot \url{https://github.com/features/copilot}} Codeium,\footnote{Codeium \url{https://codeium.com/}} CodeWhisperer,\footnote{CodeWhisperer \url{https://aws.amazon.com/codewhisperer/}} \textit{etc.}.
As AI-supported programming tools grow and evolve, this reshapes the culture and context of software development~\cite{Brady2023}. This growth also influences the design of AI-based tools with a focus on improving their accuracy, optimizing memory usage, and enhancing other competitive parameters (\eg, ~\cite{gunasekar2023textbooks,pires2023one,zhou2023dataset,geiping2023cramming}). 

However, while these technical aspects are important, it is equally crucial to consider their impact on the end-users. Therefore, besides improving technical performance, academic discourse should address implications for user satisfaction and overall user experience. Research revealed that AI supports programmers' productivity in some ways, but, on the other hand, developers now tend to spend more time reviewing code than writing it~\cite{mozannar2023reading}. Since the time spent on code comprehension is directly related to \CR --- the more readable the code is, the less time the developer needs to understand it --- we can optimize the programmers' workflow by providing suggestions from an LLM that align with their notion of \CR. In academic research, \CR is usually defined as a subjective, mostly implicit human judgment of how easy the code is to understand~\cite{posnett2011,buse2008,scalabrino2016improving}.

Currently, within the field of artificial intelligence and machine learning, researchers are working towards enhancing the suitability of LLMs for specific domains~\cite{ling2023domain}, for instance, natural language processing~\cite{duval2021breaking} and medical applications~\cite{karabacak2023embracing}, rather than focusing on meeting user expectations. LLM suitability for specific domains is enhanced through fine-tuning, where the model's weights and parameters are modified using task-specific data. However, this fine-tuning process also offers the potential for aligning LLMs with humans. Approaches like Reinforcement Learning from Human Feedback~\cite{lambert2022illustrating}, Preference Elicitation and Preference Learning~\cite{blum2004preference}, In-Context Learning~\cite{dong2023survey} can be harnessed to adjust models based on human preferences, resulting in improved usability, accessibility, and user satisfaction. 

Such an alignment in terms of \CR requires a profound understanding of developers' perceptions of what constitutes readable code. This involves having a substantial labeled dataset of code that captures human preferences. Alternatively, to mitigate the resource-intensive nature of data labeling, one can employ a model trained on a smaller dataset to predict labels for unlabeled data~\cite{settles2009active}. For this purpose, existing predictive models of \CR~\cite{posnett2011,scalabrino2018comprehensive,dorn2012general,mi2022} may potentially be used, despite being initially created as a foundation for static analysis tools aimed at helping developers make their code more readable. 




Our research aims to assess whether the existing \CR models evaluate the readability of AI-generated code similarly to how developers do. For this purpose, we (a) created an AI-generated dataset of code, (b) evaluated AI-generated code with existing \CR models, (c) understood what aspects of code developers rely on when assessing \CR, (d) conducted a human evaluation of AI-generated code in terms of readability, (e) compared the assessments made by \CR models and developers. This makes our work not only innovative but also significant in the context of modern software engineering.

Overall, our research represents a pioneering effort in directing attention towards \CR from a developer's perspective and assessing whether the existing \CR models appropriately represent their standpoint within the context of AI-generated code. Using the Repertory Grid technique, a tool from cognitive psychology, we identified 12 dimensions that developers consider essential for \CR. We then generated a dataset of readability-labeled Java code snippets, on which we reevaluated existing \CR models. 

Our research indicates a weak correlation between current \CR models and developer evaluations, pointing to a significant gap in these models' ability to reflect developers' perspectives on \CR. It underscores the need for developing more accurate \CR metrics and models.

The contributions of this research can be summarized as follows:

\textbf{Understanding AI Output Readability} --- to our best knowledge, we were the first to evaluate the readability of current AI-generated code using established \CR models and human raters and compare these evaluations. Our findings indicate that when the generated solution is meaningful, concise, and adheres to style guidelines, it tends to be considered readable both by \CR models and human assessors. This underscores the effectiveness of AI systems, to some extent, in producing code that is easy to read by humans, but also emphasizes the importance of ongoing development and refinement of LLMs to improve both the readability and functionality of AI-generated code.

\textbf{Human-centered representation of Readability} --- our research contributes to the field by identifying 12 specific aspects of code that influence \CR, as perceived by developers with varying experience levels. This discovery extends the network of factors considered in \CR research, offering a foundation for creating and maintaining code that is more easily understood by humans, thereby enhancing developer productivity and collaboration.

\textbf{Advancements in Readability evaluation} --- our finding of the low level of alignment between existing \CR models and human judgments and the low agreement level between human raters within the context of AI-generated code adds a novel dimension to the current understanding of \CR assessments. It underscores the need for further research on readability-related code aspects and the human-aligned \CR models. 
    
\textbf{Future research directions} --- to encourage further exploration and innovation, we have made our study's code and data available.\footnote{For access to the code and data used in our research, visit the link: \url{https://zenodo.org/records/10550937}.} 

\section{BACKGROUND} \label{sec:background}

\subsection{Code-fluent LLMs as Coding Assistants}
The development of code-fluent LLMs as coding assistants has fundamentally transformed the coding experience. Several research studies were conducted to examine how humans and AI interact in-depth and to understand how LLMs influence programmer behavior during coding activities~\cite{mozannar2023reading, liang2023understanding, vaithilingam2022expectation}.

The researchers found that developers spend approximately 50\% of coding time in interaction with the LLM, with 35\% dedicated to double-checking suggestions since the generated code is limited in meeting both functional and non-functional requirements. Developers also struggle to comprehend the outputs of LLM. 

The studies highlighted the importance of aligning models with human requirements, reducing coders' time and mental effort to comprehend AI coding assistants' suggestions. Developers need to quickly comprehend the code proposed by an AI coding assistant before integrating it into a project and implementing any changes. A critical aspect of this process is what is commonly referred to as \CR --- the ease with which developers can read and understand code. In this notion, \CR forms a perceived barrier to comprehension that developers must overcome to efficiently work with code~\cite{posnett2011,buse2008,scalabrino2018comprehensive}.

\subsection{Alignment of Code-fluent LLMs}
Addressing developers' requirements regarding the readability of model-suggested code may involve various fine-tuning methods. Specifically, in addition to the fine-tuning process itself, when the developer modifies the model's weights and parameters, contrastive~\cite{le2020contrastive} and reinforcement~\cite{lambert2022illustrating} learning are valuable tools for this purpose. The implementation of these methods encompasses the definition of a learning objective --- information about what output is ``desirable'' and what is not. 


Creating a new dataset for adjusting models requires substantial human and computational resources. It involves the collection of vast amounts of data, preliminary testing, processing, and analysis, followed by multiple refinement iterations to ensure optimal performance. To minimize annotation efforts of data-demanding machine-learning tasks, several approaches are suggested. These include reusing pre-trained models on new tasks (\ie transfer learning~\cite{hosna2022transfer}), using both labeled and unlabeled data for training (\ie semi-supervised learning~\cite{van2020survey}), using a model that was trained on a smaller subset of data to predict the labels of the unlabeled data (\ie active learning~\cite{settles2009active}). 

Our research focused on exploring the potential of existing \CR models~\cite{posnett2011,dorn2012general,scalabrino2018comprehensive,scalabrino2016improving,mi2018improving,mi2022} to be learning objectives to guide the process of fine-tuning. These \CR models, designed to represent and predict human evaluations, can aid in the fine-tuning process of LLMs, essentially modeling the readability of the training data. 

By examining whether existing \CR models accurately reflect human evaluations, we can potentially repurpose them as benchmarks for the adaptation and tuning of LLMs, reducing the time, effort, and resources typically required for this task.

\begin{table*}[h]
\centering
\begin{tabular}{lcccc}
\hline
\textbf{Authors} & \textbf{\# of Java Snippets} & \textbf{\# of Annotators} & \textbf{\# of Features} & \textbf{Accuracy\textasteriskcentered} \\\hline
Buse and Weimer~\cite{buse2008}& 100 & 120 & 25 & 0.80 \\
Posnett et al.~\cite{posnett2011} & \multicolumn{2}{c}{Taken from Buse and Weimer} & 3 & 0.83 \\
Dorn~\cite{dorn2012general} & 120 & 2789 & 7 & F1 $\sim$0.81 \\
Scalabrino et al.~\cite{scalabrino2018comprehensive} & 200 & 9 & 104 & 0.85 \\
Mi et al.~\cite{mi2022} & \multicolumn{2}{c}{Taken from previous works} & Unknown & 0.853 \\\hline
\end{tabular}

\begin{tablenotes}
\centering
\footnotesize
    \item \textasteriskcentered The percentage of code snippets correctly classified as readable/unreadable.    
\end{tablenotes}
\caption{Some critical characteristics of SOTA Models of \CR.}

\label{tab:sota_tab}
\vspace{-20pt}
\end{table*}

\vspace{-0.4cm}
\subsection{Models of \CR}

\textbf{Buse's Model.} Buse and Weimer pioneered a method to quantify \CR using machine learning~\cite{buse2008}. They crafted a binary classifier model based on code features they hypothesized would influence \CR. This model was trained using readability assessments from 120 computer science students on 100 Java code snippets, rated on a 1 to 5 scale. Despite the subjective nature of readability and only moderate inter-rater agreement, their model successfully predicted readability and aligned with established software quality metrics.

The model utilized metrics such as code structure, complexity, and documentation, which were processed into vectors for machine learning. The classifier demonstrated an 80\% accuracy rate and a Pearson correlation of .63 with the mean human readability scores, suggesting its effectiveness in \CR prediction. A principal components analysis further revealed that six out of the 25 features accounted for most of the variance in readability, highlighting the potential for a more streamlined model.




\textbf{Posnett's Model.} Posnett, Hindle, and Devanbu introduced a Simpler Model of \CR based on only three features, which surpassed the performance of Buse and Weimer's earlier model~\cite{posnett2011}. Leveraging the same dataset as Buse and Weimer, they applied a forward stepwise refinement for feature selection, in contrast to the original backward stepwise approach. Conversely, Posnett's team manually incorporated features, assessing their consequent impact on the model's quality. The selection of features in their model was driven by the authors' intuition and familiarity with Halstead's software science metrics.

Their findings indicate that a less complex model, incorporating Halstead volume, token entropy, and line count, can more effectively predict \CR. However, this study did not overcome the limitations of Buse and Weimer's lack of operationalization of \CR, as it relied on the same dataset and did not assess the code evaluators' expertise and understanding of \CR.

\textbf{Dorn's Model.} Concerns about the generalizability of previous Java-based readability models led Dorn to develop a General Software Readability Model, which broadened the scope to multiple programming languages~\cite{dorn2012general}. Dorn's dataset encompassed 360 code snippets from open-source projects in Java, Python, and C++ with CUDA, varying from 10 to 50 lines. Over 5,000 participants rated the snippets' readability on a scale from 1 to 5.

Dorn's approach extended beyond syntactic analysis to include structural patterns, visual perception, alignment, and natural language elements, transformed into numerical vectors. Using logistic regression, Dorn identified seven key features that most significantly correlated with human readability judgments, achieving a Spearman correlation of .72. This model also surpassed the retrained Buse and Weimer model by 5\% in F-measure, underscoring the value of incorporating a broader range of code characteristics into readability assessments.





\textbf{Scalabrino's Model.} Scalabrino, Linares‐Vásquez, and Oliveto expanded on \CR research by proposing a Comprehensive Model that integrates syntactic, visual, structural, and textual elements of code~\cite{scalabrino2018comprehensive}. They curated a dataset from 200 methods across four Java projects, rated on readability by 9 CS students.

Merging their dataset with those from Buse and Weimer and Dorn, they computed features from all preceding models plus new textual features. Using logistic regression, they crafted a binary \CR classifier with 104 features. This model surpassed the previous ones, showing at least a 6.2\% increase in accuracy, underscoring the benefit of textual alongside structural and syntactic features in \CR prediction.




\textbf{Mi's Model.} Mi, Hao, Ou, and Ma introduced a deep-learning-based \CR model that transcends the need for manual feature engineering by leveraging visual, semantic, and structural code representations~\cite{mi2022}. Their approach utilized an image recognition model for visual features, word embeddings for semantic understanding, and a character matrix for structural attributes, culminating in a composite neural network.

This model, with a threefold approach to feature extraction, fed into a neural network classifier, outperformed traditional machine learning models on a combined dataset from Buse and Weimer, Dorn, and Scalabrino. With an average accuracy of 85.3\%, it marked an improvement ranging from 3.5\% to 13.8\%, showcasing the potential of deep learning in automating \CR evaluation.

\vspace{-0.5cm}

\subsection{Suitability for Fine-Tuning of \CR Models}
Among the five described models, four are acknowledged as state-of-the-art: Posnett's, Dorn's, Scalabrino's, and Mi's. While pioneering, the model introduced by Buse and Weimer has subsequently been outperformed by the advancements of its counterparts. The remaining four models of \CR do, however, exhibit certain shortcomings in their suitability for the alignment of LLMs.

To begin with, these models are more code-centric as opposed to human-centric. They primarily rely on either hand-crafted or hidden features. This can result in misinterpretation or neglect of important \CR aspects, leading to sub-optimal decisions during LLMs' adjustment. The hand-crafted feature method, applied in Posnett's, Dorn's, and Scalabrino's models, may not fully encapsulate all the crucial factors affecting \CR since the features selected for these models are solely based on researchers' intuition. In contrast, Mi's deep learning model does not explicate specific features it uses for the assessments. Therefore, it is possible that none of the abovementioned methods adequately encapsulates \CR aspects developers perceive as important. The neglect of human-centric considerations in these models might lead to a mismatch between the AI’s understanding of \CR and that of real-world developers.

Secondly, there seems to be a lack of a standardized or universally accepted understanding of \CR among participants. The ratings of \CR in the datasets of all four studies are based on subjective assessments using a Likert scale, and in most cases, these ratings were provided by computer science students who may not fully represent the diversity of the developer population. This raises concerns about the generalizability of the results. LLMs adjusted using biased data may end up reinforcing the wrong patterns, rendering the outputs less readable for developers.


Therefore, while the aforementioned \CR models can be potentially used for fine-tuning purposes, their limitations must be addressed for effective and accurate optimization of LLMs. It is essential to check if these models are robust and comprehensive enough and adequately capturing the diverse readability notions from different developers' perspectives. 

\subsection{Present Research}

To check whether the evaluations made by existing Java \CR models are aligned with the evaluations of \CR made by Java developers in the context of AI-generated code, the present study aims to answer four research questions:

\textbf{RQ 1.} How readable is Java code produced by a code-fluent LLM, according to existing \CR models' evaluations? 

\textbf{RQ 2.} What are the aspects of AI-generated code that influence its readability according to Java developers with varying levels of experience?

\textbf{RQ 3.} How do Java developers with varying experience levels assess the readability of the code produced by a code-fluent LLM?

\textbf{RQ 4.} Is there a correlation between humans' and models' evaluations of \CR in AI-generated Java code?

To answer these questions, we conducted a study consisting of three main parts: (a) creating an AI-generated dataset of code, (b) conducting a structured interview applying the Repertory Grid technique~\cite{manual}, and (c) a large-scale labeling survey. \change{
The study was conducted in line with our company's ethical standards, adhering to the values and guidelines outlined in the ICC/ESOMAR International Code~\cite{iccesomar}. The study's schematics overview can be found in the online appendix.}
\section{AI-GENERATED DATASET OF JAVA CODE} \label{subsec:dataset}

The crucial aspect of our research is a set of AI-generated code snippets needed for the interview and labeling survey. We deliberately created code snippets from scratch rather than employing segments from existing open-source projects since our primary objective was to enhance human-AI interaction. Consequently, our focus centers on evaluating the readability of the current outputs generated by LLMs and examining how well these human evaluations align with assessments made by existing \CR models.

\subsection{Data Collection}
We selected Code Golf challenges \footnote{Code Golf game \url{https://code.golf/}} to generate AI-based code snippets, aligning with the brevity required by the \CR models under study. After extracting all 95 tasks, we noted that specific tasks existed in two versions --- long and short. In our research, we retained the longer versions since solving them required a more substantial piece of code. Furthermore, we excluded tasks that required Unicode characters, like chess symbols or emojis. This selection process resulted in a final set of 64 programming problems that were well-suited for prompting the LLM. Consequently, we employed ChatGPT 3.5 Turbo \change{(due to the timing of the study)} to generate Java language solutions for these tasks. 

Using ChatGPT 3.5 Turbo, we generated Java solutions with three prompt variations: Basic, Intermediate, and Expert, following best practice guidelines for specificity and clarity~\cite{google-best-practices}. \textit{Basic}: to write a straightforward Java program for a given task within 50 lines, without explanations; \textit{Intermediate}: to develop a Java program adhering to formatting and best practices in efficiency, readability, and maintainability, limited to 50 lines; \textit{Expert}: to create an advanced Java program following best programming practices and formatting conventions, capped at 50 lines, using an example for guidance.

To ensure that the brevity of generated snippets did not impact their meaningfulness, two authors of current research looked through them, validating if the generations were acceptable for our research. Then, snippets were tested on executability. Executable snippets were further assessed for adherence to a length limit of 50 lines, as defined by the examined \CR models. 

The outcome of these tests was 120 Java code snippets of solutions for 49 tasks ---  15 tasks were excluded from the original tasks' list because the model could not produce meaningful and executable solutions within 50 lines of code. Therefore, in the final set of snippets, some solved the same task but were generated by prompts of different levels.

\subsection{Data Analysis}

To assess the readability of AI-generated Java code, we utilized existing predictive models for \CR: Posnett's~\cite{posnett2011}, Dorn's~\cite{dorn2012general}, Scalabrino's~\cite{scalabrino2018comprehensive}, and Mi's~\cite{mi2022}. Using these models, we labeled each snippet based on the readability score as assessed by each model. We also assigned the snippet a label of the overall readability according to all four models --- ``Readable'' (R) if three or more models deemed the snippet readable; ``Unreadable'' (U) if the opposite was true; and ``Not Defined'' (N) if the snippet was considered both readable and unreadable by two models each. 

Posnett's model was presented in the form of a formula~\cite{posnett2011} that we applied to calculate the probability of snippets being readable. Dorn's model was shared with us by the author along with the script used for calculating the readability of snippets. Scalabrino's model has an open-source script for determining readability,\footnote{Scalabrino et al., Experimental material, raw data, and readability tool: \url{https://dibt.unimol.it/report/readability}} which we utilized in our analysis. Finally, Mi's model was employed from their open GitHub repository.\footnote{Mi et al., \CR model: \url{https://github.com/swy0601/readability-Features}}

To assess the impact of the prompting strategy on snippet readability, we conducted pairwise comparisons between groups using the $\chi^2$ test with each readability model.

\subsection{Findings}

The abovementioned analysis was aimed to answer \textbf{RQ 1. How readable is Java code produced by a code-fluent LLM, according to existing Code Readability models' evaluations?}

In our research, when AI generates Java code that is both executable and concise, we found that it is mostly considered readable according to the existing \CR models (see ~\Cref{tab:readability-results}). The results also highlight the significant variance across different \CR models' classification of the same AI-generated Java code snippets. Of particular note is the high contrast between Posnett's model, which found most snippets readable, and Dorn's, Scalabrino's, and Mi's models, which found a more significant proportion of snippets unreadable. 

Given this significant variability in evaluations, the question of how closely these models align with human assessments of \CR becomes particularly salient.

We also noted instances where AI-generated code was unexecutable, excessively lengthy, and unreadable. This highlights that while impressive strides have been made in the readability of AI-generated code, there is still significant room for improvement. 

\begin{table*}[ht]
\centering
\begin{tabular}{@{}lcccc@{}}
\toprule
 & \textbf{Posnett's} & \textbf{Dorn's} & \textbf{Scalabrino's} & \textbf{Mi's} \\
\midrule
\# of Readable & 118 (97.52\%) & 77 (64.89\%) & 73 (60.83\%) & 82 (68.42\%) \\
\# of Unreadable & 2 (2.48\%) & 43 (35.11\%) & 47 (39.17\%) & 38 (31.58\%) \\
\bottomrule
\end{tabular}
\smallskip
\caption{Readability of AI-Generated Snippets According to Existing \CR Models.}
\label{tab:readability-results}
\vspace{-20pt}
\end{table*}

Additionally, the results of the $\chi^2$ test revealed that the prompts did not have a statistically significant effect on readability for either group (p > .05). Given this finding, as well as the fact that solutions frequently had observable distinctions based on prompt types, we opted to treat them as a unified set of snippets for subsequent research purposes. We justify this approach because at no point in our research participants could see the entire set of snippets. This means they could not encounter all possible solutions for a single task. If, by random assignment, such a situation did occur during the survey stage, our survey metrics would account for it, and it would not affect the validity of our results.

\section{INTERVIEW} \label{sec:interview}

Previous studies of \CR models relied on the creation of their datasets utilizing subjective readability ratings without a consistent definition. To address this limitation, in our research, we set out to provide the proxy of standardized understanding of \CR by providing annotators readability-related code aspects to evaluate prior to \CR in the labeling process, therefore ``priming'' them on a unified implicit understanding of \CR.

To investigate the aspects influencing \CR from the developers' perspective, we conducted one-hour long online interviews, employing the Repertory Grid technique~\cite{kelly2003psychology}. Details about this technique, including how it facilitated the collection of important \CR aspects and their qualitative analysis, are provided further in this section.

\subsection{Sample}

We interviewed 15 Java Developers of different experience levels --- Beginner, Intermediate, and Advanced --- with five participants in each group. As gratitude for participating in the research, external developers were offered a 100 USD Amazon eGift Card or declined any compensation.

The participants were picked from those developers who previously participated in our research and gave permission for future contact, as well as internal colleagues. They were invited to participate in a screening survey where they shared the years they spent in professional programming, including internships. Participants were also asked to self-evaluate their proficiency in Java. 

Based on their responses, we categorized participants into three groups that effectively represent our sample for research purposes (\change{we treated outlier cases \eg ``Advanced with less than 3 years of experience'' as impossible and used them as a basis for excluding participants from further analysis}):

\textit{Beginners:} developers with less than one year of coding experience who considered themselves Intermediate and those with less than five years of experience who rated themselves as Beginners. 
    
\textit{Intermediate:} developers with more than one year but less than ten years of coding experience who self-identified as Intermediate. We also included developers who considered themselves Beginners with over five years of experience and those who defined themselves as Advanced with less than three years of experience.

\textit{Advanced:} developers with more than three years of programming experience who viewed themselves as Advanced and those with over ten years of professional experience who classified themselves as Intermediate.

\subsection{Repertory Grid Technique}

The Repertory Grid technique~\cite{manual} was proposed by George Kelly in 1955, and it centers on his Personal Construct Theory~\cite{kelly2003psychology}. According to this theory, people perceive the world through bipolar cognitive constructs like Happy/Sad or Good/Bad. These constructs play a key role in shaping an individual's understanding and interpretation of the world around them, influencing their behavior. In essence, the Repertory Grid technique is a tool employed to elicit these individual's implicit constructs by utilizing their language and cognitive frameworks, ensuring that research findings remain grounded in the person's reality.

The technique builds on interviews for constructs' elicitation and further questionnaires to validate the obtained constructs~\cite{EDWARDS2009785}. This procedure initiates by determining a specific area of investigation. In our case, this area is \CR. It continues by identifying relevant elements within the defined area. For our study, the elements are AI-generated code snippets. The next step requires the individual to compare and contrast these elements in triads to extract constructs that, from their perspective, are essential for distinguishing different elements of a defined area --- in our case, determining code snippets as readable or unreadable. For example, such code characteristics as ``Nesting'' might be a \CR construct represented by poles ``Code is flat and linear''  and ``Code is overly nested''. This characteristic may be elicited when three code snippets with various levels of nesting are compared. 

Once the constructs are gathered, the individuals are asked to rate each element (\ie code snippet) against all identified constructs. This is usually done using a Likert scale. In the context of our research, this phase was presented as the labeling process described below in~\Cref{sec:survey}. 

The collected ratings are subsequently subject to an analysis aiming to delineate the individual's perception of the area. In our research, this understanding was used to assess the degree of alignment between existing \CR models and developers.

\vspace{-0.4cm}
\subsection{Data Collection}

We selected random snippets from the initial set of 120, described in the ~\Cref{subsec:dataset}, and grouped them into ten unique triplets. The triplets comprised all potential combinations of the ``Readable'' (R), ``Unreadable'' (U), and ``Not Defined'' snippets --- RRR, RRU, RRN, RUU, RUN, RNN, UUU, UUN, UNN, and NNN. These triplets served as illustrative material for the Repertory Grid interview, providing diverse examples of \CR.

During the session, participants first provided background information about their education, language skills, and programming tools, establishing a context for their insights. In the training phase, they evaluated three code snippets for \CR, selecting and discussing the most similar pair and differentiating the third. This process revealed their \CR constructs. In the main phase, participants engaged in multiple rounds of evaluating triplets of code snippets, with the order of triplets randomized to prevent sequence bias. Due to the one-hour limit, most participants reviewed four triplets before ending with a debriefing for questions and additional comments.

\subsection{Data Analysis}

During the data analysis, we compiled a comprehensive yet concise list of key aspects influencing \CR from all the constructs obtained during the interviews. 

The initial collection of bipolar constructs representing \CR aspects was presented independently to three experts: a Java Developer who had no prior knowledge of the study, a Data Analyst who is one of the authors but wasn't involved in the interviews, and a Cognitive Psychologist who is one of the authors and conducted several interviews during the study. Their task was to categorize these aspects into logical and distinct groups and then name these groups of constructs, drawing on their specialized knowledge and expertise. Afterward, the experts engaged in synchronous and asynchronous discussions to sort and group constructs into coherent clusters until a unanimous decision was reached. 

\subsection{Findings}

The Repertory Grid interview was utilized to answer \textbf{RQ 2. What are the aspects of AI-generated code that influence its readability according to Java developers with varying levels of experience?}

This technique facilitated the identification of overall 123 bipolar \CR constructs. For descriptive statistics, see~\Cref{tab:descr_cons_tab}.

\begin{table}[ht]
\centering
\begin{tabular}{@{}lcccc@{}}
\toprule
& \multicolumn{2}{c}{\textbf{\# of Triplets Seen}} & \multicolumn{2}{c}{\textbf{\# of Constructs Elicited}} \\
\cmidrule(lr){2-3} \cmidrule(lr){4-5}
\textbf{Group} & \textbf{Mean} & \textbf{SD} & \textbf{Mean} & \textbf{SD} \\
\midrule
Beginner & 4.2 & 0.8 & 7.8 & 1.3 \\
Intermediate & 5.2 & 1.6 & 8.0 & 1.2 \\
Advanced & 5.0 & 2.0 & 8.8 & 2.6 \\
\bottomrule
\end{tabular}
\smallskip
\caption{Descriptive statistics for elicited \CR constructs.}
\label{tab:descr_cons_tab}
\vspace{-20pt}
\end{table}

Subsequently, the list of constructs was subjected to a consolidation process led by three experts who identified 15, 11, and 8 clusters of constructs associated with \CR, respectively. The next step involved a synchronous discussion among the experts, where 26 distinct clusters, some of which were similar in meaning to each other, were merged into a final set of 12 unique, comprehensive clusters (see~\Cref{tab:aspect-mentions}).


\begin{table}[h]
\centering
\begin{tabular}{@{}lcccc@{}}
\hline
\textbf{Code Aspect} & \textbf{Overall} & \textbf{Beg.} & \textbf{Int.} & \textbf{Adv.} \\
\hline
\rowcolor{gray}
Code Structure & 30 & 12 & 6 & 12 \\
Code Style & 15 & 4 & 7 & 4 \\
\rowcolor{gray}
Naming & 14 & 5 & 3 & 6 \\
Sufficient Contextual Info & 11 & 4 & 5 & 2 \\
\rowcolor{gray}
Familiar Code Patterns & 9 & 2 & 3 & 4 \\
Reading Flow & 8 & 3 & 3 & 2 \\
\rowcolor{gray}
Inline Actions & 8 & 3 & 2 & 3 \\
Understandable Goal & 7 & 2 & 3 & 2 \\
\rowcolor{gray}
Code Length & 7 & 1 & 5 & 1 \\
Nesting & 6 & 1 & 1 & 4 \\
\rowcolor{gray}
Visual Organization & 4 & 1 & 1 & 2 \\
Magic Numbers Usage & 3 & 0 & 1 & 2 \\
\hline
\end{tabular}
\smallskip
\caption{Readability-Related Code Aspects and the Number of Their Mentions Among Participants.}
\label{tab:aspect-mentions}
\vspace{-20pt}
\end{table}

Later, we converted this set into a rating list by assigning a bipolar description for each characteristic.
These descriptions were chosen in synchronous expert discussions as the most relevant and representative pair of readability-related code aspects obtained from the participants.

The constructs described above collectively highlight that from the developers' point of view \CR requires a balance between being informative and concise, structured yet simple, and conforming to a known pattern while being visually appealing and user-friendly. For the distribution of mentions of specific clusters in the interview, refer to ~\Cref{tab:aspect-mentions}.

We employed these characteristics of code to create leading questions in a further labeling survey. This way, raters can assess readability consistently and share a common understanding of which aspects of code to consider when evaluating \CR.
\section{LABELING SURVEY}\label{sec:survey}

We initiated a labeling survey to craft a carefully labeled dataset of AI-generated Java code snippets, reflecting their \CR based on human perceptions of the concept. This labeled dataset was then employed to evaluate the existing \CR models as learning objectives for LLM fine-tuning.

\subsection{Sample}

The sample was gathered by sending the survey link to the list of those Java programmers who previously participated in our surveys and research and gave permission for future contact. None of them has participated in the interview stage of the current study. As a thank you, participants were given the opportunity to enter a draw for one of five 100 USD Amazon eGift Cards or an equivalent-value company product pack.

The online data labeling process engaged 390 Java programmers of different experience levels, each contributing at least one snippet evaluation. On average, participants completed seven assessments.

All of the participants are proficient in the Java programming language. Among them, 78\% know more than three programming languages. 
Most participants are Developers, Programmers, and Software Engineers --- 362 of 390. Additionally, Software Architects and DevOps Engineers participated, as well as programmers with other various job roles. 


To determine the participants' proficiency, we gathered information about their years of programming experience and their self-assessed Java expertise. \change{The distribution of programming experience among participants is as follows: 27 individuals with less than 1 year, 60 with 1–2 years, 55 with 3–4 years, 44 with 5–6 years, 42 with 7–8 years, 20 with 9–10 years, and 142 with more than 10 years.} 36 participants self-assessed as Beginners, 162 as Intermediate, and 192 as Advanced developers.



\subsection{Materials}

In the labeling survey, we utilized materials gathered during the previous stages of the research --- the set of AI-generated Java code snippets and the rating list of \CR aspects.
The primary purpose of this list in our research was to provide unified guidance to the labelers during the assessment of \CR, ensuring that the labelers evaluated code consistently and with a focus on key readability-related aspects.

\subsection{Data Collection}

On the greeting page of the survey, participants gave their consent and professional background information. After that, they were presented with a randomly assigned Java code snippet from the previously generated dataset of snippets (see ~\Cref{subsec:dataset}). Participants evaluated \CR of the snippet using the list of bipolar characteristics with a five-point scale measuring how much the code leans to the readable or unreadable pole of characteristics. Furthermore, respondents answered a single-choice question to identify whether the code was readable. 


\subsection{Data Analysis}

The next step in investigating the correlation between human \CR evaluations and predictions generated by \CR models involved assigning readability labels to each code snippet based on human assessments. For that, the results from the survey were employed. 

We calculated Krippendorff's alpha to assess the agreement of readability evaluations among labelers since this metric can handle cases of various numbers of raters~\cite{feng2015mistakes}. After that, code snippets that received a readable rating from more than 50\% of raters were marked as readable and the remaining as unreadable. As a result of this process, a fifth readability label was added to each code snippet in the dataset, and the four labels were assigned based on evaluations by the \CR models.

Following this labeling process, we assessed each model's alignment with human ratings. This was done using a Matthews correlation coefficient as an established metric to measure the correlation between binary data. By this, we identified the model that showed the most substantial level of correlation with human ratings of \CR.

\subsection{Findings}

We conducted the labeling survey to answer RQs 3 and 4 about agreement in assessments between developers and existing \CR models. 

\begin{figure*}[h]
    \centering
    \includegraphics[width=0.7\textwidth]{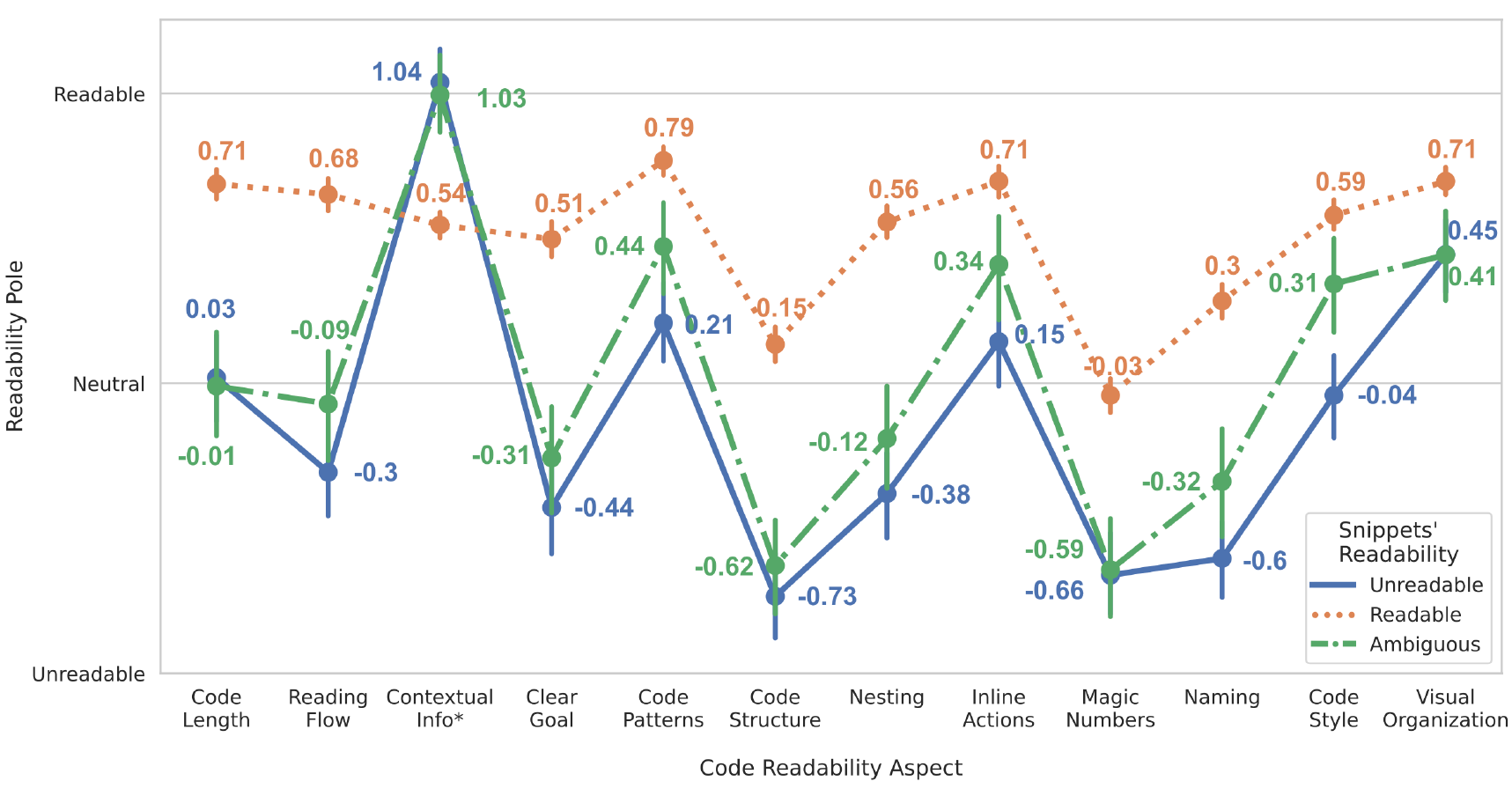}
    \begin{minipage}{\textwidth}
        \centering
        \footnotesize
        \textcolor{darkgray}{\textasteriskcentered This characteristic is ``Readable'' when Neutral and ``Unreadable'' at the extremes}
    \end{minipage}
    \caption{Visualisation of mean score for each \CR aspect.}
    \label{fig:CRval}
    \Description{The graph features mean scores for each code readability aspect for three groups of snippets: Unreadable, Readable, and Ambiguous, as reflected in the legend. 
    The x-axis represents code readability aspects, with 'Code Length' on the left and 'Visual Organization' on the right. Aspects considered here are Code Length, Reading Flow, Contextual Info, Clear Goal, Code Patterns, Code Structure, Nesting, Inline Actions, Magic Numbers, Naming, Code Style, Visual Organization. 
    The y-axis represents mean scores, categorized as 'Unreadable' at the bottom, 'Neutral' in the middle, and 'Readable' at the top. 
    The Readable group consistently achieves higher scores, closely approaching the Readable pole. In contrast, the Unreadable and Ambiguous groups tend to have lower scores.
    A pattern emerges in the Unreadable and Ambiguous groups, with all aspects showing similar trends. However, there are some points where the Ambiguous group's score also appears close to the Readable group. Namely Code Patterns, Inline Actions, Code Style, and Visual Organization.
    Another noteworthy aspect is 'Contextual Info,' initially considered readable when neutral and unreadable at the extremes. Consequently, the scores for the Unreadable and Ambiguous groups in this aspect appear higher than those for the Readable group.}
    \end{figure*}

\textbf{RQ 3. How do Java developers with varying experience levels assess the readability of code produced by a code-fluent LLM?}

The agreement level on \CR assessments between humans was found to be low in general ($\alpha$ = .14) as well as within different levels of expertise --- $\alpha$ = .04 for Beginners, $\alpha$ = .1 for Intermediate, $\alpha$ = .16 for Advanced. There is a possibility that the difficulty of assigning the readability label to the code, reflected in these results, is dictated by the subjectivity of \CR and its aspects. We discuss these results in more detail further in the paper (see ~\Cref{sec:discussion}).

To assign a particular readability label to each code snippet, given the low level of agreement among labelers, we labeled snippets according to the majority of votes. If there were no definitive majority of votes (\eg five readable labels and six --- unreadable), the snippets, which we refer to as ``Ambiguous'' snippets, were excluded from further analysis since if human raters could not agree on their readability, there is no point in measuring compliance of the models with this evaluation. As a result of this process, we categorized 109 code snippets --- 93 as ``Readable'' and 16 as ``Unreadable'', discarding 11 code snippets from further analysis.

\change{In~\Cref{fig:CRval}, mean scores for \CR aspects are shown for each snippet group. The x-axis represents readability aspects, and the y-axis represents scores on a five-point scale measuring how much the code leans to the readable or unreadable pole of characteristics, as asked in the survey.}

\textbf{RQ 4. Is there a correlation between humans' and models' evaluations of \CR in AI-generated Java code?}

The results indicate that Scalabrino's model~\cite{scalabrino2018comprehensive} exhibits a moderate correlation with human evaluations (MCC = 0.325). In contrast, others have a weaker correlation --- MCC = 0.033 for Mi’s model, MCC = -0.036 for Posnett’s, and MCC = -0.038 for Dorn’s --- which makes them not very suitable for serving as the learning objective in the fine-tuning of code-fluent LLMs, particularly regarding the readability of their output.

\section{DISCUSSION} \label{sec:discussion}


The main goal of our research was to check whether evaluations of \CR made by existing models are aligned with the evaluations of \CR made by Java developers by answering the following RQs. 


\paragraph{\textbf{RQ 1. How readable is Java code produced by a code-fluent LLM, according to existing \CR models' evaluations?}}

The readability of Java code snippets produced by a code-fluent LLM was first evaluated by each of the existing \CR models, including Posnett's~\cite{posnett2011}, Dorn's~\cite{dorn2012general}, Scalabrino's~\cite{scalabrino2018comprehensive}, and Mi's~\cite{mi2022}. 

Overall, we found that AI-generated Java code, when it is both executable and concise, is mostly readable according to the existing \CR models. However, the results yielded significant variability across evaluations from these models. For instance, according to Posnett's model, only 2.48\% of snippets in the dataset were unreadable. Contrarily, the assessment from Dorn's model was considerably stricter, identifying 35.11\% of the snippets as unreadable. Similarly, Scalabrino's model and Mi's model found nearly 39.17\% and 31.58\% of snippets unreadable, respectively.

These findings underscore not only the effectiveness of AI systems in producing human-like code but also the importance of continuous development and refinement to enhance both the readability and functionality of AI-generated code. Further research into why discrepancies in models' evaluations occur --- what definitions, standards, or aspects of \CR are employed in each model --- is needed. 

Moreover, these discrepancies suggest that attempting to formulate one all-encompassing definition for \CR may not fully capture the complex landscape of the realm as determined by diverse models. These findings supported our research on how well these models align with the human notion of \CR.

\observation{While AI systems can produce human-like and easy-to-read code, there is still room for improvement. Also, notable differences in readability assessments of AI-generated code among various \CR models highlight the need for investigating the extent to which these models represent human judgments and what factors contribute to them.}

\paragraph{\textbf{RQ 2. What are the aspects of AI-generated code that influence its readability according to Java developers with varying levels of experience?}}

To answer this RQ, we employed the Repertory Grid technique to extract 12 distinct code aspects that influence \CR perceptions among Java developers with varying expertise levels. 

We empirically confirmed that, from a developer's perspective, \CR  depends on balancing brevity and understanding. The code's purpose plays a crucial role, with readable code clearly expressing its goal, while unreadable code may obscure its intention. Readable code follows a logical structure, with actions separated into distinct lines and named constants for clarity. On the other hand, unreadable code might be excessively nested, contain multiple actions in a single line, and use undefined ``magic numbers''. Readable code often adheres to style guides and maintains a visually balanced distribution of color blocks, potentially related to syntax highlighting. In contrast, unreadable code may suffer from poor formatting and distracting color block patterns. 

The readability-related code aspects identified in our study align with prior research. 

For instance, Fakhoury and colleagues examined over 500 code readability enhancement commits and found that if they explicitly aimed at improving readability, they often involved changes in Complexity, Documentation, and Size metrics. Their study also noted that the most substantial readability improvements were seen in import statements, code formatting and style, and the reduction of ``magic numbers''.

In their fMRI study, Peitek et al.~\cite{9402005} examined 41 complexity metrics and their impact on program comprehension, finding that both the textual length and vocabulary size of code increase cognitive load and working memory demand for programmers.


There are also similarities between obtained \CR characteristics in the current research and those from previous models. Our results show that developers perceive code as readable based not only on structural characteristics of the code as assumed by earlier models proposed by Buse and Weimer ~\cite{buse2008}, and Posnett et al. \cite{posnett2011}, rather on the combination of structural characteristics with visual, textual, and linguistic features as proposed by later models \cite{dorn2012general, scalabrino2018comprehensive,mi2022}. For instance, Visual Organization from current research can be viewed as similar to Visual features from Dorn's and Mi's models. Magic Numbers Usage and Naming echo Natural Language Features from Dorn's model and  Identifier Terms in Dictionary from Scalabrino et al.'s research. 

At the same time, unlike in previous studies, we have also identified characteristics of \CR that are connected with the meaning of the code that a person tries to extract to comprehend the code --- Understandable Goal, Sufficient Contextual Info and Familiar Code Patterns. 

These findings again emphasize that readability is a complex, individual concept tied to how developers mentally represent the code and make sense of it.

\observation{Tools and strategies used to improve \CR should focus on balancing brevity with understanding, maintaining logical structure, adhering to style guides, and providing clearly expressed code purposes while accounting that no one universal standard of \CR might exist.}

\paragraph{\textbf{RQ 3. How do Java developers with varying experience levels assess the readability of code produced by a code-fluent LLM?}}

To address this RQ, we employed the list of readability-related code aspects collected during the Repertory Grid stage of the current research. We used this list in a labeling survey where participants evaluated the readability of 120 AI-generated Java methods designed for solving Code Golf game tasks.

The labeling survey results show that, despite the efforts to provide guidance during \CR assessment to ensure consistent evaluation of snippets based on key readability-related aspects, individuals tend to assess readability subjectively, and their evaluations do not always consistently align with one another. These findings are in line with prior \CR studies, where human annotators exhibited imperfect agreement, having a correlation around .5 with the mean readability score~\cite{buse2008,dorn2012general}. 

 
\observation{The inherent subjectivity in readability assessments by developers, even when guided by key readability-related aspects, highlights the need for more generalizable, operationalized, and validated definitions of \CR. The aim of future research could be to explore the relationships between \CR characteristics obtained from developers and their weights in one's decision whether the code is readable or not.}

\paragraph{\textbf{RQ 4. Is there a correlation between humans' and models' evaluations of \CR in AI-generated Java code?}}

Our findings reveal variations in the correlation between existing \CR models and human evaluations. Notably, Scalabrino's model~\cite{scalabrino2018comprehensive} exhibits a moderate correlation with the human assessments. Other models, such as Posnett's~\cite{posnett2011}, Dorn's~\cite{dorn2012general}, and Mi's~\cite{mi2022} demonstrate substantially weaker correlations. 

\change{The low MCC scores stem from the balanced nature of the metric. Despite Posnett's model classifying 97.52\% (118/120) and humans classifying 85.32\% (93/109) of snippets as readable, indicating an expected moderate to high correlation, the observed correlation is weak (MCC = -0.036). This weakness arises because the metric considers both true positives and true negatives, and in cases of significant class imbalance (Posnett's model classifying only two samples as unreadable), each miss in the smaller class carries more weight. As Posnett's model and our participants didn't match in any unreadable cases, it resulted in near-zero correlation.}

These findings are consistent with prior research on the relationship between human judgments and metrics of code understandability, which includes readability. Scalabrino et al. demonstrated that metrics typically used for effort estimation and associated with understandability, like cyclomatic complexity, often have little to no correlation with actual understandability~\cite{8115654}.

This disparity in correlation values has implications for the utility of these \CR models as learning objectives in the fine-tuning process of code-fluent LLMs to improve the readability of their generated outputs. The moderate correlation observed with Scalabrino's model suggests that it might have some potential for enhancing the readability of code generated by LLMs. Still, a more robust and precise model for guiding the adjustment process is yet to be developed and might be built utilizing the insights from current research regarding code aspects crucial for \CR.

\observation{The fairly low correlation between existing models and human evaluations of \CR implies that these models could be refined or complemented with human-centered aspects to guide better the process of adjusting code-fluent LLMs output aimed at enhancing the readability of AI-generated code.}

\subsection{Overall Discussion}

Our study explored the potential use of existing \CR models for adjusting LLMs to programmers' needs. The aim was to investigate whether these models could significantly correlate with human ratings. Such a finding could save resources in future research since having a robust model of human perspective could eliminate the need for creating extensive datasets for fine-tuning. 

However, our findings revealed that none of the existing \CR models significantly correlated with human ratings. Additionally, we observed that humans did not consistently agree on \CR evaluations, even after receiving leading assessments of readability-related code aspects. This highlights the challenge posed by the subjectivity of readability-related factors and \CR itself when aligning LLMs. If readability is such a subjective metric, it may necessitate an individualized model's adjustment in each case or breaking this metric down to several less complex ones.

Nonetheless, there is potential for certain aspects of code that impact readability to be universally applicable to most programmers or specific programmer groups. This underscores the need for further research in this area. Future studies should focus on validating code aspects related to \CR, developing models that incorporate these elements, and creating a comprehensive, accurately labeled dataset accordingly. With these components in place, achieving alignment of LLMs with the human notion of \CR becomes possible.

\section{Threats to Validity}\label{sec:ttv}

\textbf{Construct Validity} refers to the accuracy with which a study's measurements capture and represent the concepts under research. In our study it involves determining whether the constructs identified during interviews and employed in the labeling survey adequately represent how developers conceptualize \CR and whether the measurements of \CR truly reflect it. We used the well-regarded Repertory Grid technique to ensure construct validity for operationalizing \CR. \change{Recognizing potential oversights due to a limited number of interviewees and their perspectives and expert viewpoints, we included individuals with diverse experience levels to enhance our operationalization.}
Further validation of \CR related constructs was beyond our research scope, which focused on establishing a consistent understanding of \CR among survey participants.


We acknowledge that the concept of \CR, when reduced to binary terms, overlooks the complexity and the subjective nature inherent in the evaluation of code quality. However, the purpose of this research was to examine the correlation between the assessments of \CR by developers and the evaluations derived from models that utilize binary labels. To maintain methodological consistency and to facilitate a direct comparison, the binary labeling approach was adopted for the survey. Future research may consider a more granular approach to better capture the multifaceted aspects of \CR.

We addressed the potential inaccuracy of \CR models by sourcing them meticulously from original resources, like papers, communications with authors, and cited websites with the models. This ensured the models' measurements aligned with their authentic formulations, reinforcing the reliability of our findings.


\textbf{Internal Validity} ensures that a study identifies outcomes without being affected by external factors. In the interview phase, interviewers received peer-reviewed training on how to apply the Repertory Grid technique consistently and unbiasedly, ensuring the internal validity of the elicited \CR aspects. 

In the labeling survey phase, we maintained internal validity through a web-based platform, presenting AI-generated code snippets to all participants in a unified way with unified instructions. Although these snippets were displayed in a Dark color scheme, unfamiliar and inaccessible to some and potentially impacting their \CR assessments, our robust sample size of 390 participants helps mitigate this limitation and weaken other possible minor intervening factors. 

We randomized snippet presentations in both the interview and survey phases to counter potential order effects. 

\textbf{External Validity} refers to the extent to which our findings can be generalized to other contexts. We acknowledge that our findings focus primarily on \CR of AI-generated Java code snippets perceived by Java programmers. Hence, this study does not address how these results extend to \CR in other programming languages. 

Moreover, we acknowledge that Code Golf tasks (a deliberate methodological choice for generating AI-based code snippets aimed at examining LLM capabilities in a controlled environment, focusing on the clarity and conciseness of code) may not encompass the typical complexity found in larger codebases.  Extending our research by including a diverse range of programming tasks that mirror the real-world scenarios in which LLMs are commonly employed would provide a more comprehensive understanding of LLMs' potential in everyday coding practices.
\section{CONCLUSION} \label{sec:conclusion}

In this research, our primary objective was to investigate whether existing \CR models could function as a proxy for the human evaluations of \CR. We aimed to explore the possibility of using these \CR models to align the outputs of LLMs with the expectations and needs of programmers, all in the pursuit of enhancing programmers' productivity when working with AI coding assistants.

To achieve this, we created a set of 120 AI-generated Java code snippets and took a human-centered approach to define the notion of \CR in this code, considering the diverse perspectives of developers. Utilizing the Repertory Grid technique, we identified 12 distinct code aspects that affect \CR. These aspects gave us a framework for understanding \CR in AI-generated Java code, which was instrumental in creating a dataset for evaluating existing \CR models. To gather this dataset, we conducted a survey involving 390 Java programmers who assessed AI-generated Java code snippets in terms of the abovementioned aspects of \CR.

Our findings indicate that when AI generates executable and concise code, it tends to be readable. Additionally, developers' \CR assessments are based on their implicit notions about \CR and thus involve subjectivity, leading to differing opinions among human raters. Furthermore, we observed that only Scalabrino's \CR model moderately correlated with human readability ratings. This suggests that the existing models may not be well-suited for guiding the process of aligning AI-generated code with human notions of \CR.

Further research aimed at enhancing programmer efficiency might, drawing on the data from this study, focus on developing a \CR model or creating a comprehensive labeled dataset for LLM adjustment, thereby improving the alignment of AI-generated code with human readability standards.




\bibliographystyle{ACM-Reference-Format}
\bibliography{refs}


\end{document}